\documentclass[iop,revtex4,twocolumn,apj]{emulateapj}
\usepackage{apjfonts} 

\shorttitle{CGM of SMGs} 
\shortauthors{Fu et al.}
\journalinfo{}
\submitted{}

\newcommand{\hers}{{\it Herschel}}

\newcommand{\kms}{{km s$^{-1}$}}
\newcommand{\cmsq}{{cm$^{-2}$}}
\newcommand{\ms}{$M_{\rm star}$}
\newcommand{\mg}{$M_{\rm gas}$}
\newcommand{\mh}{$M_{\rm halo}$}

\newcommand{\msun}{$M_{\odot}$}
\newcommand{\msunyr}{${\rm M}_{\odot}~{\rm yr}^{-1}$}
\newcommand{\lsun}{$L_{\odot}$}
\newcommand{\um}{$\mu$m}
\newcommand{\uJy}{$\mu$Jy}

\newcommand{\sqdeg}{deg$^2$}
\newcommand{\Ha}{H$\alpha$}
\newcommand{\lya}{Ly$\alpha$}
\newcommand{\lyb}{Ly$\beta$}
\newcommand{\HI}{H\,{\sc i}}
\newcommand{\OI}{O\,{\sc i}}
\newcommand{\CII}{C\,{\sc ii}}
\newcommand{\CIV}{C\,{\sc iv}}
\newcommand{\SiII}{Si\,{\sc ii}}
\newcommand{\SiIV}{Si\,{\sc iv}}
\newcommand{\AlII}{Al\,{\sc ii}}
\newcommand{\FeII}{Fe\,{\sc ii}}

\newcommand{\NII}{[N\,{\sc ii}]}
\newcommand{\MgII}{Mg\,{\sc ii}}

\newcommand{\sqps}{SMG$-$QSO pairs}
\newcommand{\sqp}{SMG$-$QSO pair}
\newcommand{\hl}{}

\begin{document}

\title{The Circumgalactic Medium of Submillimeter Galaxies. I. First Results from a Radio-Identified Sample}

\author{
Hai~Fu\altaffilmark{1}, J.~F.~Hennawi\altaffilmark{2}, J.~X.~Prochaska\altaffilmark{3}, R.~Mutel\altaffilmark{1}, C.~Casey\altaffilmark{4}, A.~Cooray\altaffilmark{5}, D.~Kere\v{s}\altaffilmark{6}, Z.-Y.~Zhang\altaffilmark{7,8}, D.~Clements\altaffilmark{9}, J.~Isbell\altaffilmark{1}, C.~Lang\altaffilmark{1}, D.~McGinnis\altaffilmark{1}, M.~J.~Micha{\l}owski\altaffilmark{7}, K.~Mooley\altaffilmark{10}, D.~Perley\altaffilmark{11}, A.~Stockton\altaffilmark{12}, and D.~Thompson\altaffilmark{13}
}
\altaffiltext{1}{Department of Physics \& Astronomy, University of Iowa, Iowa City, IA 52242}
\altaffiltext{2}{Max-Planck-Institut fur Astronomie, Heidelberg, Germany}
\altaffiltext{3}{Department of Astronomy and Astrophysics, UCO/Lick Observatory, University of California, 1156 High Street, Santa Cruz, CA 95064}
\altaffiltext{4}{Department of Astronomy, the University of Texas at Austin, 2515 Speedway Blvd, Stop C1400, Austin, TX 78712}
\altaffiltext{5}{Department of Physics and Astronomy, University of California, Irvine, CA 92697}
\altaffiltext{6}{Department of Physics, Center for Astrophysics and Space Sciences, University of California at San Diego, 9500 Gilman Drive, La Jolla, CA 92093}
\altaffiltext{7}{Institute for Astronomy, University of Edinburgh, Royal Observatory, Blackford Hill, Edinburgh EH9 3HJ, UK}
\altaffiltext{8}{ESO, Karl Schwarzschild Strasse 2, D-85748 Garching, Munich, Germany}
\altaffiltext{9}{Astrophysics Group, Imperial College London, Blackett Laboratory, Prince Consort Road, London SW7 2AZ, UK}
\altaffiltext{10}{Oxford Centre for Astrophysical Surveys, Denys Wilkinson Building, Keble Road, Oxford OX1 3RH}
\altaffiltext{11}{Dark Cosmology Centre, Niels Bohr Institute, University of Copenhagen, Juliane Maries Vej 30, 2100 K{\o}benhavn {\O}, Denmark}
\altaffiltext{12}{Institute for Astronomy, University of Hawaii, 2680 Woodlawn Drive, Honolulu, HI 96822}
\altaffiltext{13}{Large Binocular Telescope Observatory, University of Arizona, 933 N. Cherry Ave, Tucson, AZ 85721}

\begin{abstract}
We present the first results from an on-going survey to characterize the circumgalactic medium (CGM) of the massive high-redshift galaxies detected as submillimeter galaxies (SMGs). We constructed a parent sample of 163 SMG-QSO pairs with separations less than \mbox{$\sim$36\arcsec} by cross-matching far-infrared-selected galaxies from \mbox{\it Herschel} with spectroscopically confirmed QSOs. The \mbox{\it Herschel} sources were selected to match the properties of SMGs. We determined the sub-arcsecond positions of six \mbox{\it Herschel} sources with the Very Large Array and obtained secure redshift identification for three of those with near-infrared spectroscopy. The QSO sightlines probe transverse proper distances of 112, 157, and 198~kpc at foreground redshifts of 2.043, 2.515, and 2.184, respectively, which are comparable to the virial radius of the \mbox{$\sim10^{13}$~\msun} halos expected to host SMGs. High-quality absorption-line spectroscopy of the QSOs reveals systematically strong \mbox{\HI} \mbox{\lya} absorption around all three SMGs, with rest-frame equivalent widths of \mbox{$\sim2-3$}~\AA. However, none of the three absorbers exhibits compelling evidence for optically thick \mbox{\HI} gas or metal absorption, in contrast to the dominance of strong neutral absorbers in the CGM of luminous \mbox{$z \sim 2$} QSOs. The low covering factor of optically thick \mbox{\HI} gas around SMGs tentatively indicates that SMGs may not have as prominent cool gas reservoirs in their halos as the co-eval QSOs and that they may inhabit less massive halos than previously thought.
\end{abstract}


\section{Introduction} \label{sec:intro}

\begin{deluxetable*}{rrrccc crrcc}
\tablewidth{0pt}
\tablecaption{VLA-observed \hers\ Sources
\label{tab:vlaphoto}}
\tablehead{
\colhead{Pair Name} & \colhead{RA$_{250}$} & \colhead{Dec$_{250}$} & \colhead{$S_{250}$} & \colhead{$S_{350}$} & \colhead{$S_{500}$} & \colhead{Int Time} & \colhead{RA$_{\rm 6 GHz}$} & \colhead{Dec$_{\rm 6 GHz}$} & \colhead{$S^{\rm peak}_{\rm 6 GHz}$} & \colhead{$S^{\rm int}_{\rm 6 GHz}$} \\
\colhead{} & \colhead{(deg)} & \colhead{(deg)} & \colhead{(mJy)} & \colhead{(mJy)} &  \colhead{(mJy)} &  \colhead{(min)} & \colhead{(deg)} & \colhead{(deg)} & \colhead{(\uJy/bm)} & \colhead{(\uJy)} \\
\colhead{(1)} & \colhead{(2)} & \colhead{(3)} & \colhead{(4)} & \colhead{(5)} & \colhead{(6)} & \colhead{(7)} &  \colhead{(8)} & \colhead{(9)} & \colhead{(10)} & \colhead{(11)} 
}
\startdata
HeLMS 0015$+$0404  &  3.9286&$+$4.0715 & 61.7$\pm$6.0 & 67.8$\pm$5.7 & 54.3$\pm$7.2&26.5&  3.93038&$+$4.07262  & 69.4$\pm$6.7 & 72.4$\pm$13.4\\
HeLMS 0041$-$0410  & 10.3541&$-$4.1679 & 80.8$\pm$6.2 & 83.0$\pm$6.5 & 43.6$\pm$7.0&15.8& \nodata & \nodata & $<$35.1& \nodata  \\
L6-XMM 0223$-$0605 & 35.8056&$-$6.0860 & 33.8$\pm$2.3 & 40.3$\pm$2.5 & 24.2$\pm$3.5&69.7& \nodata & \nodata & $<$14.4& \nodata  \\
G09 0918$-$0039    &139.6159&$-$0.6644 & 41.1$\pm$6.9 & 49.7$\pm$8.1 & 31.2$\pm$9.1&18.9& \nodata & \nodata & $<$17.7& \nodata  \\
G09 0920$+$0024    &140.2475&$+$0.4049 & 35.0$\pm$7.0 & 51.6$\pm$8.1 & 32.0$\pm$8.9&22.9& \nodata & \nodata & $<$18.6& \nodata  \\ 
NGP 1313$+$2924    &198.4530&$+$29.4126& 59.7$\pm$5.6 & 78.5$\pm$6.6 & 53.6$\pm$7.8&17.2& \nodata & \nodata & $<$42.3& \nodata  \\
NGP 1330$+$2540    &202.5866&$+$25.6749& 49.2$\pm$5.8 & 54.3$\pm$6.4 & 29.3$\pm$7.8&18.0& \nodata & \nodata & $<$15.3& \nodata  \\ 
NGP 1333$+$2357    &203.3743&$+$23.9592& 30.4$\pm$5.4 & 31.8$\pm$6.4 & 29.1$\pm$7.5&55.2&203.37514&$+$23.95909 & 20.0$\pm$3.0 & 31.0$\pm$8.1 \\
NGP 1335$+$2805    &203.9409&$+$28.0986& 41.7$\pm$5.5 & 49.8$\pm$6.4 & 38.2$\pm$7.7&27.6&203.94249&$+$28.09750 & 38.3$\pm$4.5 & 51.4$\pm$11.0\\
G15 1413$+$0058    &213.4580&$+$0.9725 & 46.6$\pm$6.4 & 52.8$\pm$7.7 & 36.1$\pm$8.5&16.2&213.45743&$+$0.97321  & 37.3$\pm$5.8 & 37.3$\pm$11.6\\
G15 1435$+$0110    &218.9043&$+$1.1682 & 63.0$\pm$6.7 & 63.8$\pm$8.0 & 56.6$\pm$8.8& 9.2&218.90494&$+$1.16958  & 75.6$\pm$7.3 & 89.1$\pm$16.0\\
G15 1450$+$0026    &222.6773&$+$0.4351 & 47.5$\pm$6.9 & 47.9$\pm$8.1 & 29.1$\pm$8.9&16.3& \nodata & \nodata & $<$18.0& \nodata \\
L6-FLS 1712$+$6001 &258.0352&$+$60.0281& 32.3$\pm$2.2 & 34.0$\pm$2.4 & 22.4$\pm$3.6&30.1&258.03111&$+$60.02722 & 10.8$\pm$2.1 & 10.8$\pm$4.2 \\
 & & & & & & & 258.04010 & $+$60.02625 & 13.9$\pm$2.1 & 15.6$\pm$4.5  
\enddata
\tablecomments{
L6-FLS 1712$+$6001 has two radio counterparts (see Fig.~\ref{fig:detections}); the first line shows the nominal counterpart, which is closer to the \hers\ position although slightly fainter. 
Columns (2-6) list the \hers\ 250~\um\ positions and the photometry at 250, 350, and 500~\um.
Column (7) is the total VLA on-source integration time.
Columns (8-9) list the positions of the radio counterparts.
Columns (10-11) are the peak flux density in \uJy/bm and the integrated flux density in \uJy, both of which are derived by fitting an elliptical Gaussian model to the source. The uncertainty of peak flux density is given by the rms noise in the map at the source position, while the uncertainty of the integrated flux density is estimated using the formulae provided by \citet{Hopkins03a}, which includes the 1\% uncertainty in the VLA flux-density scale at 6~GHz \citep{Perley13}.  
}
\end{deluxetable*}

The first milli-Jansky-level submillimeter surveys discovered a population of distant submillimeter-bright galaxies (SMGs), namely, unresolved sources with 850~\um\ flux density ($S_{850}$) greater than 3$-$5~mJy \citep{Smail97,Barger98,Hughes98,Eales99}. The SMGs selected at wavelengths between 850~\um\ and 1~mm are intense starbursts (SFR $\gtrsim 500$~\msunyr) at a median redshift of $z \sim 2.5$ \citep{Chapman05,Wardlow11,Yun12,Smolcic12}. The intense star formation is dust-enshrouded so that the SMGs radiate most of their bolometric luminosity in the far-infrared (IR). The observed molecular and stellar emission indicates that they are massive gas-rich galaxies \citep[\hl{\mbox{$M_{\rm gas} \sim M_{\rm star} \sim 10^{11}$\,\msun}}; e.g., ][]{Michalowski10a,Hainline11,Bothwell13}, but the typical halo mass of SMGs remains uncertain, with estimates ranging from $10^{12}$ to $10^{13}$\,\msun. Two lines of evidence suggest that SMGs may inhabit dark matter halos as massive as $\sim10^{13}$~\msun: (1) their strong clustering strength estimated from either the angular two-point correlation function \citep[e.g.,][]{Scott06,Weis09} or the cross-correlation function between SMGs and other high-redshift galaxies \citep[e.g.,][]{Hickox12}, and (2) their high stellar mass and the \ms$-$\mh\ relation from abundance matching \citep[\mh~$= 6\times10^{12}$~\msun\ for \ms~$= 10^{11}$~\msun\ at $z = 2$; e.g.,][]{Behroozi10}. However, because source blending due to the large beams of single-dish (sub)millimeter telescopes may have significantly elevated clustering strength \citep{Cowley16} and the stellar mass estimates remain uncertain within an order of magnitude \citep[e.g.,][]{Hainline11,Michalowski12, Targett12}, it is possible that a typical SMG may inhabit much less massive halos ($\sim$10$^{12}$\,\msun).

SMGs are absent in the local universe and it is commonly thought that they have evolved into the massive ellipticals today \citep[e.g.,][]{Blain04a,Toft14}. To understand the evolution of SMGs, it is imperative to know how long the observed intense star formation would last. For $10^{13}$~\msun\ halos at $z = 2.5$, the average baryonic accretion rate from the mass growth rate of dark matter haloes is $\dot{M}_{\rm gas} \equiv 0.18 \times \dot{M}_{\rm halo} \simeq 1.4\times10^3$~\msunyr\ \citep{Neistein08,Bouche10}. In such massive halos, it is expected that most of the baryons will be shock-heated to the virial temperature of the halo ($\sim10^7$~K) so that only a small fraction of the accreted gas can actually cool and accrete onto galaxies \citep[e.g.,][]{Keres05, Dekel06}. Therefore, the ongoing gas accretion is unlikely to sustain the extreme SFRs. Without a comparable gas supply rate, the SFR would decline with an $e$-folding timescale of only $\sim$200~Myr \citep[2\mg/SFR; e.g.,][]{Greve05, Tacconi08, Ivison11, Bothwell13, Fu13}. At such a rate, the SMGs would become red sequence galaxies in only a Gyr or 5 $e$-folding times\footnote{This is the time it would take to decrease the specific SFR (SFR/\ms) from $\sim10^{-9}$~Gyr$^{-1}$ for the SMGs at the observed epoch to $\sim10^{-11}$~Gyr$^{-1}$ for the red sequence at $z \sim 2$ \citep{Brammer09}.}. Such a short transitional time of a significant high-redshift star-forming population might help explain the rapid build-up of the massive end of the red sequence at $z > 1$ \citep[e.g.,][]{Ilbert13}. Starbursts thus provide an alternative mechanism to the QSO-mode feedback \citep[e.g.][]{Silk98,Di-Matteo05} to form red and dead galaxies. Note that both mechanisms still require feedback from radio jets \citep[i.e., the maintenance mode; e.g.,][]{Fabian12,Heckman14} to prevent the hot gaseous halo from cooling. On the other hand, if SMGs were in $10^{12}$~\msun\ halos, the intense star formation is also unsustainable because the gas accretion rate is only $\sim$110~\msunyr\ at $z = 2.5$. 

However, could there be enough cool gas in the circumgalactic medium (CGM) around SMGs to fuel a prolonged starburst phase \citep[e.g. see the simulation of][]{Narayanan15}? The CGM of co-eval QSOs may give us a hint, because they inhabit comparably massive ($\sim10^{12.6}\,M_{\odot}$) halos \citep{White12}. Contrary to the expected dominance of virialized X-ray plasma, absorption line spectroscopy of a statistical sample of $z \sim 2$ projected QSO pairs reveals the prevalence of cool ($T \sim 10^4$~K), metal-enriched ($Z \geq 0.1 Z_\odot$), and optically thick \lya\ absorbers ($N_{\rm HI} \geq 10^{17.2}$~\cmsq) extending to at least the expected virial radius of 160~kpc (the ``QSO Probing QSO'' [QPQ] project: \citealt{Hennawi06a,Hennawi07,Prochaska13,Prochaska13a}). The high observed covering factor of the cool CGM gas ($\gtrsim 60\%$) in $\sim10^{12.6}\,M_{\odot}$ halos
has been compared to predictions from numerical simulations. While several studies found that they cannot reproduce the high covering factor around QSOs (\citealt{Fumagalli14, Faucher-Giguere15}, but see \citealt{Rahmati15}), it has been argued that efficient star-formation-driven winds from accreted satellite galaxies that interact with cosmological filaments are required to increase the \HI\ covering factor to the observed level, which is only resolved in the highest resolution cosmological zoom simulations \citep{Faucher-Giguere16}.

These high-resolution simulations predict that the covering factor is roughly independent of SFR, decreases with redshift, and has relatively large halo-halo variations. In the relevant halo mass range, $\sim 3\times 10^{12}-10^{13}$\,\msun, simulations show \HI\ covering factors of $\sim 30-80\%$ with little mass dependence at $z \sim 2$ \hl{\mbox{\citep{Faucher-Giguere16}}}. Given the estimated halo masses for QSOs and SMGs, one could expect similar covering factors if these are determined primarily by the interplay between gas infall and star-formation-driven outflows. To test this, we exploit QSO absorption line spectroscopy to probe the CGM of SMGs. We first present the data sets and the method we used to select projected \sqps\ in \S~\ref{sec:sample}. We then describe our followup observations in \S~\ref{sec:obs}, including radio interferometer imaging, near-infrared spectroscopy, and optical spectroscopy. We present our analysis and results in \S~\ref{sec:result}, including a comparison between the covering factor of optically thick gas around SMGs and that of $z \sim 2$ QSOs. We summarize the results and conclude in \S~\ref{sec:summary}. Throughout we adopt a $\Lambda$CDM cosmology with $\Omega_{\rm m}=0.27$, $\Omega_\Lambda=0.73$ and $H_0$ = 70 km~s$^{-1}$~Mpc$^{-1}$.  

\section{Selection of Projected SMG$-$QSO Pairs} \label{sec:sample}

Because both high-redshift QSOs and SMGs have low number counts, we need large samples of both to come up with a sizable sample of projected \sqps\ with small angular separations. We compiled 464,866 spectroscopically confirmed QSOs from various surveys: primarily, the Sloan Digital Sky Survey \citep[SDSS;][]{Alam15}, the 2dF QSO Redshift Survey \citep[2QZ;][]{Croom04}, the AGN and Galaxy Evolution Survey \citep[AGES;][]{Kochanek12}, and the MMT Hectospec Redshift Survey of 24 {$\mu$}m Sources in the Spitzer First Look Survey \citep{Papovich06}. Since we select the foreground galaxies that are likely at $z > 2$ (see the next paragraph), we keep only the 102,472 QSOs at $z_{\rm QSO} > 2.5$. The average surface density of these background QSOs is $\sim$10\,deg$^{-2}$. 

To select the foreground SMGs, we combined source catalogs from a number of wide-area extragalactic surveys carried out by the \hers\footnote{{\it Herschel} is an ESA space observatory with science instruments provided by European-led Principal Investigator consortia and with important participation from NASA.} Space Observatory \citep{Pilbratt10}: the \hers\ Multi-tiered Extragalactic Survey \citep[HerMES, 95 \sqdeg;][]{Oliver12,Wang14a}, the \hers\ Astrophysical Terahertz Large Area Survey \citep[H-ATLAS, 600 \sqdeg;][]{Eales10,Valiante16}, the \hers\ Large Mode Survey \citep[HeLMS, 301 \sqdeg;][; Clarke et al. in prep.]{Oliver12,Asboth16,Nayyeri16}, the \hers\ Stripe 82 Survey \citep[HerS, 79 \sqdeg;][]{Viero14}. All of these surveys used SPIRE \citep[Spectral and Photometric Imaging Receiver;][]{Griffin10} to image the sky at 250, 350, and 500~\um\ and the combined xID250\footnote{250, 350 and 500~\um\ fluxes were all extracted at source positions detected on the 250~\um\ map \citep[e.g.,][]{Roseboom10,Rigby11}.} catalog contain 1,586,047 sources covering a total of 767 \sqdeg\ (HerS and HeLMS fields overlap by 10\,\sqdeg). 

However, most \hers\ sources are not SMGs; they are, instead, less luminous dusty star-forming galaxies at lower redshifts \citep[$z < 2$;][]{Casey12a,Casey14}. To select \hers\ sources that are likely SMGs, we chose only the subsample that satisfies the following criteria: (1) flux densities peak at 350~\um\ ($S_{250} < S_{350}$ and $S_{500} < S_{350}$; i.e., ``350\um\ peakers''), (2) $S_{500} > 20$~mJy, and (3) $>$3$\sigma$ detections in all three SPIRE bands. Criterion 1 is essentially a photometric redshift selection because emission from dusts at $T = 35$~K would peak at 350~\um\ if redshifted to $z \sim 2.5$. \hl{This is confirmed by the blind carbon monoxide (CO \mbox{$J=1-0$}) survey of a subsample of the \mbox{\it brightest} \mbox{350~\um} peakers (\mbox{$S_{350} \geq 115$~mJy}), which has shown a strikingly similar redshift distribution as \mbox{850~\um}-selected SMGs \mbox{\citep[$z_{\rm CO} = 2.5\pm0.8$;][]{Harris12}}. But note that most of these bright sources are strongly lensed and they do not overlap with our sample.} Criterion 2 is introduced to ensure that the Rayleigh-Jeans extrapolation would give $S_{850} > 3$~mJy, the classic definition of an SMG, given a typical power-law slope of 3.5 for a modified blackbody with a frequency-dependent absorption cross section ($\kappa \propto \nu^{1.5}$). Criterion 3 ensures that all of the sources we considered are statistically significant. This is necessary because the image depth varies substantially from field to field, ranging from $\sigma_{500} = 15$\,mJy\,beam$^{-1}$ for the large HeLMS and HerS fields \citep[confusion noise included;][]{Oliver12,Viero14} to confusion limited with $\sigma_{500} = 6.8$\,mJy\,beam$^{-1}$ for the deeper HerMES fields \citep{Nguyen10}. \hl{Given the range of observed far-IR SEDs at \mbox{$z = 2$ from \citet{Casey12a}}, our color selection and the high threshold on the \mbox{500~\um} flux density ensure that \mbox{$\sim$95\%} of our sample would be classified as SMGs, if they were observed at 870~\um. Nevertheless, we should keep in mind that the \mbox{\hers}-selected SMGs are a subsample of SMGs and they likely cover a smaller range of dust temperatures than \mbox{870~\um}-selected SMGs \mbox{\citep[e.g.,][]{Hwang10,Magnelli12}}.} Only 70,823 \hers\ sources remained after this selection. The average surface density of 92\,deg$^{-2}$ is five times lower than the observed 870~\um\ source count above $S_{870} \gtrsim 3$~mJy \citep[$\sim$500\,deg$^{-2}$;][]{Weis09,Coppin06}. This is not surprising given that almost half of the total \hers\ area is only 10-20\% complete at $S_{500} = 20$~mJy. \hl{Note that this incompleteness in the \mbox{\it Herschel} catalogs is not a concern for compiling a sample of \mbox{\sqps}.}  
 
\begin{figure*}[!tb]
\epsscale{1.18}
\plotone{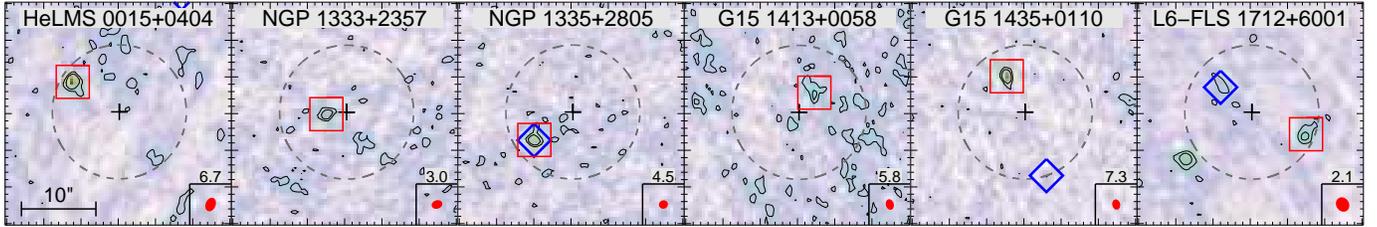}
\caption{VLA 6~GHz continuum maps for the six VLA-identified SMGs. Each image is 30\arcsec$\times$30\arcsec\ centered on the \hers\ position. The restoring beam of each map is plotted as the red ellipse at the lower right corner, above which the 1$\sigma$ noise level is labeled in units of \uJy\,beam$^{-1}$. The cross and the dashed circle indicate the \hers\ positions and the 18\arcsec\ FWHM of the 250~\um\ PSF. The red square highlights the detected radio source within the \hers\ FWHM. The blue diamond marks the optical position of the QSO, if it is within the displayed region. The contours are at ($+$2, $+$4)$\sigma$ ($\bar{\sigma} = 4.9$\,\uJy\,beam$^{-1}$). Major tickmarks are spaced in 5\arcsec\ intervals. N is up and E is left for all panels.
\label{fig:detections}} 
\end{figure*}

We identified 230 projected \sqps\ with angular separations ($\theta_{250}$)\footnote{$\theta_{250}$ is measured between the \hers\ 250\,\um\ position and the optical position of the QSO.} between 5\arcsec~$\leq \theta_{250} \leq$~36\arcsec\ by cross-matching the QSO and SMG subsamples described above. The corresponding impact parameters ($R_\bot$) are between $40 < R_\bot <$~300~kpc for $z_{\rm SMG} = 2.5$. The impact parameter is defined as the transverse proper distance at the redshift of the foreground SMG, which equals the angular diameter distance of the SMG multiplied by the angular separation on sky ($R_\bot = D_{\rm A}(z) \times \theta$). These QSO sightlines thus probe out to $\sim$1.5 virial radii of $10^{13}$~\msun\ halos ($r_{\rm vir} = 200$~kpc at $z = 2.5$). Given the positional uncertainty of the \hers\ 250~\um\ sources \hl{(\mbox{$\sigma_{\rm pos} = 3.4\arcsec$}; see \mbox{\S~\ref{sec:radiodetections}})}, the 5\arcsec\ lower limit on the angular separation is imposed as an attempt to avoid far-IR-luminous QSOs, i.e., the QSOs are the SMGs themselves. Through visual inspection of the QSO spectra, we further excluded 31 pairs whose QSOs exhibit strong broad absorption lines (BALs; this makes them unsuitable for absorption line work), have wrong redshifts, or are misclassified. Therefore, our final sample includes 199 pairs, among which 90/163 QSOs have SDSS $g \leq 21/22$ (bright enough for absorption line spectroscopy). One \hers\ source is probed by two QSOs at $\theta_{250}=$\,8.0\arcsec\ and 18.2\arcsec. No QSO probes multiple \hers\ sources within 36\arcsec, but a ``single'' \hers\ source may consists of multiple SMGs due to the large beam size; so it is possible that a single QSO could probe multiple SMGs at different impact parameters (this is indeed the case for the last source shown in Fig.~\ref{fig:detections}).

\section{Followup Observations} \label{sec:obs}

Extensive followup observations are needed to perform the absorption-line analysis of SMGs. Identifying the absorption features in the spectrum of the background QSO requires a precise redshift for the foreground SMG. But the full-width-at-half-maximum (FWHM) angular resolution of \hers\ --- 18\arcsec/25\arcsec/36\arcsec\ at 250/350/500\,\um\ --- precludes longslit spectroscopic observations with typical slit width of 1\arcsec.
To obtain more accurate positions and to identify blended sources, we exploit interferometer observations with the Karl G. Jansky Very Large Array (VLA). Furthermore, once the positions are determined with sub-arcsec accuracy, we need to determine the spectroscopic redshifts of the SMGs with near-IR spectrographs by targeting rest-frame optical lines that suffer less dust extinction than rest-frame UV lines. Finally, a high S/N optical/near-UV spectrum of the background QSO is needed to detect the UV absorption lines imprinted by the diffuse medium around the SMGs. Below we describe these observations in more details.

\subsection{SMG Identification with the VLA}

Far-IR-luminous galaxies like our \hers\ sources are expected to be luminous in the radio wavelengths, according to the IR-radio correlation \citep{Helou85,Condon92,Ivison10c}. We can thus obtain better positions for the \hers\ sources by identifying the radio counterparts with interferometers. We observed 15 \sqps\ with the VLA in the B configuration with the C-band (6~GHz) receivers (program ID: 15A-266). The sample was selected randomly from the \sqps\ with QSOs brighter than $g \leq 21$, excluding those with $\theta_{250} > $\,30\arcsec. We later realized that two of the VLA targets in the HeLMS field are likely spurious detections because of Galactic cirrus (Clarke et al. in prep.), so we excluded them in this discussion. Table~\ref{tab:vlaphoto} lists the \hers\ positions and photometry of the final VLA sample. The receivers have a total bandwidth of 4\,GHz at a central frequency of 5.9985\,GHz. The targets were selected from six different extragalactic fields. To maximize the observing efficiency, we grouped the targets with their R.A. into five scheduling blocks (SBs) of 0.8 to 3.1\,hours. A nearby unresolved calibrator was observed every $\sim$10~min. Depending on the R.A. of the targets, 3C\,48, 3C\,286, or 3C\,295 was observed for bandpass and flux-density calibration. The entire program took 8.3 hrs of VLA time. The on-source integration time ranges from 9 to 70 minutes, allocated based on the 6~GHz flux density estimated from fitting the \hers\ photometry with the SED template of a well-studied, strongly lensed SMG at $z = 2.3259$ \citep[SMM~J2135-0102, aka. the ``Eyelash'';][]{Swinbank10b}.

The observations were calibrated using the Common Astronomical Software Applications (CASA) package \citep{McMullin07}. We used the VLA pipeline to perform basic flagging and calibration. Additional flagging was performed whenever necessary by inspecting the visibility data. We used the self-calibration technique with bright sources within the primary beam to reduce the gain errors for four fields: HeLMS\,0015$+$0404, HeLMS\,0041$-$0410, L6-XMM\,0223$-$0605, and NGP\,1313$+$2924. For imaging and deconvolution, we used the standard CASA task {\sc clean} with ``natural'' weighting to achieve the best sensitivity. 
The resulted restoring beams are on average 1\farcs5$\times$1\farcs2 FWHM. The rms image noise level ranges from 2.1 to 14.1~\uJy\,beam$^{-1}$, with a mean of 6~\uJy\,beam$^{-1}$. This measured noise is consistent with thermal noise using natural weighting of the visibility data, except for maps contaminated by the sidelobes of the strong sources lying far outside the cleaned image. 

We will present the results in \S~\ref{sec:radiodetections}, but here is a summary: The VLA detected six of the 13 SMGs; For one of those, NGP~1335+2805, the QSO at $z = 2.973$ itself is responsible for the IR emission, so we excluded it from subsequent spectroscopic observations.   

\subsection{Near-IR Spectroscopy of the VLA-detected SMGs} 

\begin{figure*}[!tb]
\epsscale{0.38}
\plotone{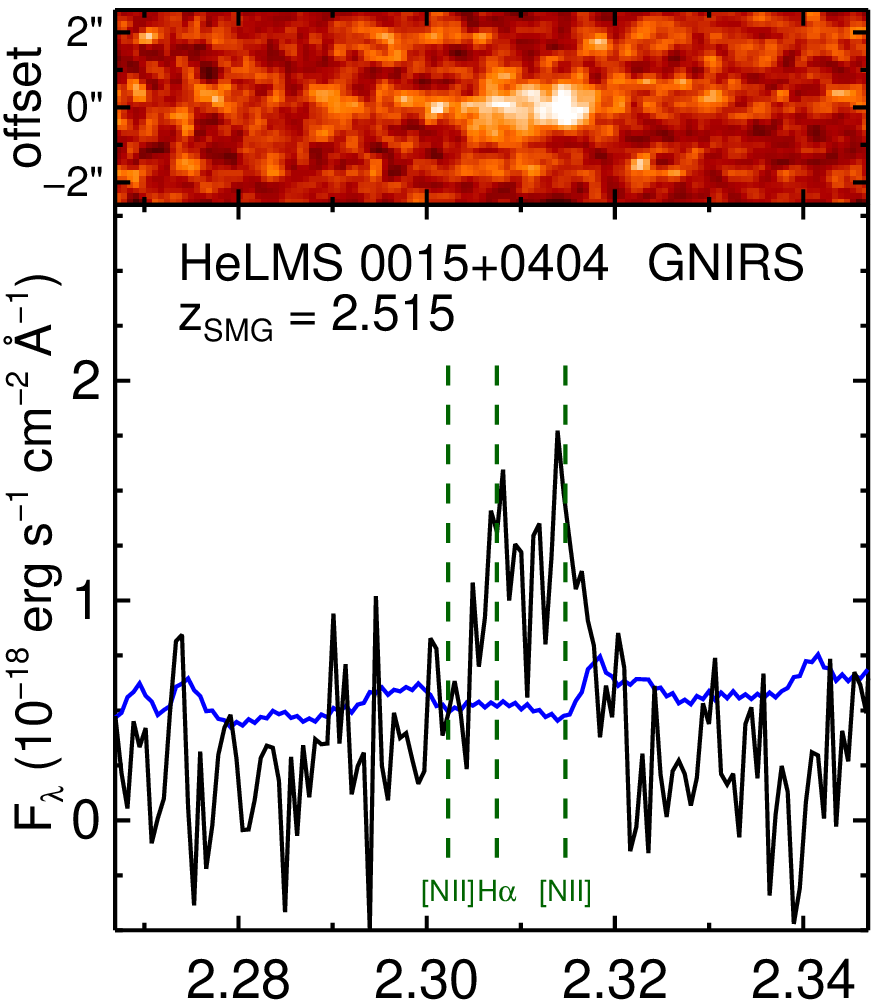}
\plotone{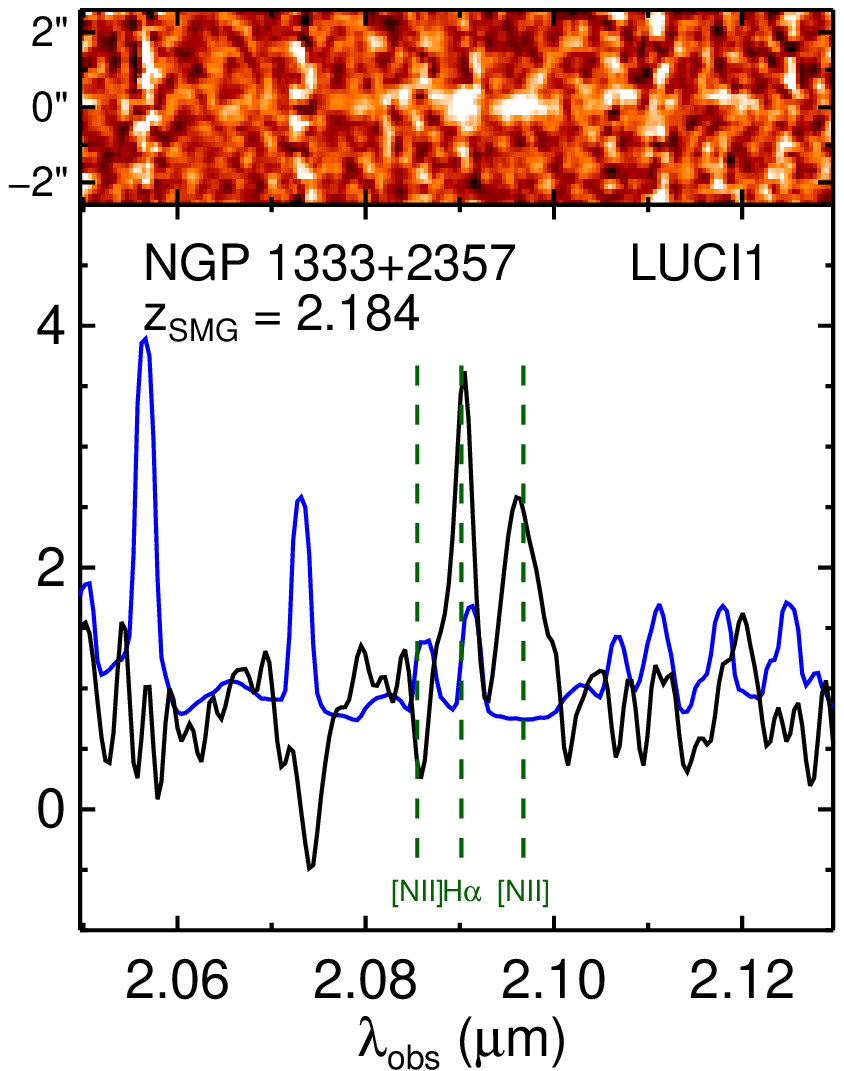}
\plotone{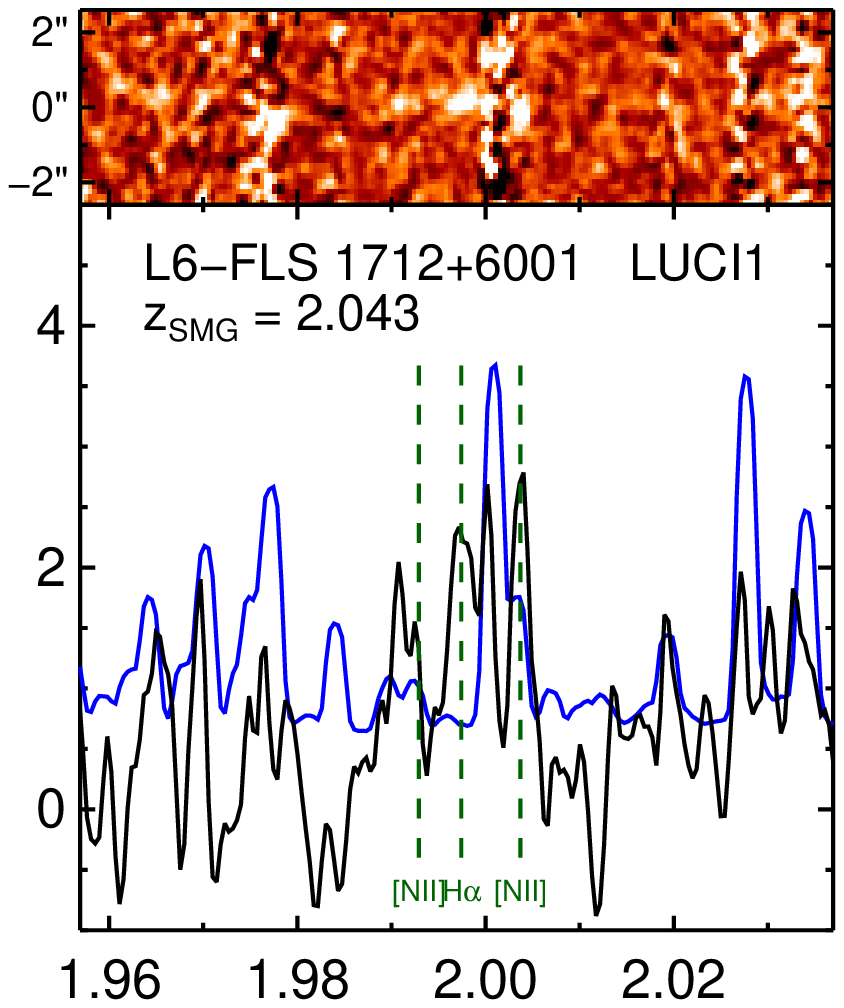}
\caption{Near-infrared spectra of the VLA-identified SMGs. The top panel shows the coadded 2D spectrum. The ordinate is the positional offset along the spatial direction, and is centered on the SMG location. The bottom panel shows the flux-calibrated 1D spectrum ({\it black}) and its 1$\sigma$ uncertainty ({\it blue}). Wavelengths affected by strong sky lines show large errors. The dashed lines indicate the redshifted \Ha\,$\lambda$6563 and \NII\,$\lambda\lambda$6548,6583 lines.
\label{fig:smgspec}} 
\end{figure*}

With the VLA positions that are accurate to $\lesssim$0.1\arcsec, we carried out a redshift survey for the five VLA-detected SMGs with near-IR spectrographs. The VLA resolved the \hers\ source of L6-FLS\,1712$+$6001 into two sources at similar brightness (Fig.~\ref{fig:detections} {\it last panel}). We chose the one closer to the \hers\ position as the nominal counterpart and have obtained a deep near-IR spectrum at that location. It is unclear whether the two sources are physically related because the redshift of the other source remains to be determined.

We observed G15\,1435$+$0110, L6-FLS\,1712$+$6001, and NGP\,1333$+$2357 with the LUCI-1 spectrograph \citep[LBT NIR-Spectroscopic Utility with Camera and Integral-Field Unit;][]{Seifert03} on the Large Binocular Telescope (LBT) on 2015 Apr 14. We used the 200 l/mm $H$+$K$ grating at $\lambda_{\rm c} = 1.93$~\um\ and the N1.8 camera (0\farcs25 per pixel) to obtain a spectral range between 1.5 and 2.3~\um. The 1\arcsec-wide 4\arcmin-long slit is centered on the VLA-determined SMG position, and it is aligned with the QSO in each pair to obtain the QSO spectrum simultaneously. The QSO spectrum is bright enough to serve as a useful reference for spatial alignment. The spectral resolution ($R$) is $\sim$940 in $H$-band and $\sim$1290 in $K$-band. We obtained 32$\times$ 120~s exposures for each target. Between exposures, we dithered along the slit among six dithering positions distributed within 40\arcsec. Atmospheric transparency varied dramatically during the night and no telluric star was observed. So we used the telluric star observation from the previous night for an approximate telluric and flux calibration.

We observed HeLMS\,0015$+$0404 and G15\,1413$+$0058 with the Gemini near-infrared spectrograph \citep[GNIRS;][]{Elias06} in the queue mode (program IDs: GN-2015B-Q-46, GN-2016A-Q-41). We used the cross-dispersing prism with the 31.7 l/mm grating and the short camera to obtain a complete spectral coverage between 0.85-2.5~\um. With the 0\farcs68 short slit, the spectral resolution is $R \sim 750$ across all orders. We obtained 14$\times$ 300~s exposures for HeLMS\,0015$+$0404 on 2015 Aug 9, Aug 12, and Oct 18, and 24$\times$ 115~s exposures for G15\,1413$+$0058 on 2016 May 20. We dithered by 3\arcsec\ along the 7\arcsec-slit between exposures.

Data reduction was carried out with a modified version of \texttt{LONGSLIT\_REDUCE} \citep{Becker09} for LUCI-1 by Fuyan Bian \citep{Bian10} and a modified version of Spextool \citep{Cushing04,Vacca03} for GNIRS by Katelyn Allers (private communication). These IDL packages carry out the standard data reduction steps for near-IR spectroscopy: flat-fielding, wavelength calibration, pairwise sky subtraction, residual sky removal, shifting and coadding, spectral extraction, telluric correction, and flux calibration. 

\hl{We identified narrow emission lines in three of the five observed SMGs, which enabled redshift measurements. We were unable to determine the redshifts of the other two sources because we detected only the continuum emission at low S/N.}

\subsection{Optical Spectroscopy of the Background QSOs}

Although previous optical spectra exist for all of the QSOs in our sample, most of them do not have the necessary wavelength coverage or sufficient S/N for absorption line analysis. Hence we obtained new optical spectra for the three background QSOs that are associated with the spectroscopically identified SMGs. We observed HeLMS\,0015$+$0404 and NGP\,1333$+$2357 on 2016 Jan 9 and L6-FLS\,1712$+$6001 on 2015 Jun 13 with the Low Resolution Imaging Spectrometer \citep[LRIS;][]{Oke95} on the Keck I telescope. We used the 1\arcsec\ longslit and the 560 Dichroic for both runs. 
For the former run, we used the 600/4000 Grism on the blue side ($R \sim 1000$ at 4,000~\AA, FWHM = 300\,\kms) and the 600/7500 grating at $\lambda_{\rm c} = 7,407$~\AA\ on the red side to cover from 3,300 to 8,700\,\AA. 
For the latter run, we used the 400/3400 Grism on the blue side ($R \sim 600$ at 4,000~\AA, FWHM = 500\,\kms) and the 400/8500 grating tilted to a central wavelength of $\lambda_{\rm c} = 8,300$~\AA\ on the red side to cover from 3,300 to 10,200\,\AA. 
The total integration time for each source ranged between 30 min and 40 min. Conditions were non-photometric for both nights. \hl{We obtained useful spectra for all three background QSOs.}

We reduced the raw data with \texttt{XIDL}, an IDL data reduction package for a number of spectrographs written by two of us (JXP and JFH). The pipeline follows the standard data reduction steps and reduces the blue and red channels separately. It begins by subtracting a super bias from the raw CCD frames, tracing the slit profiles using flat fields, and deriving the 2D wavelength solution for each slit using the arcs. Then it flat-fields each slit and rejects cosmic-rays, identifies objects automatically in the slit, and builds the bspline super sky model without rectification \citep{Kelson03}. After subtracting the super sky model, it performs optimal 1D extraction based on the spatial profile of the QSO \citep{Horne86}. Finally, it removes instrument flexure using isolated sky lines, performs heliocentric correction, and does flux calibration.

\begin{figure*}[!tb]
\epsscale{1.02}
\plotone{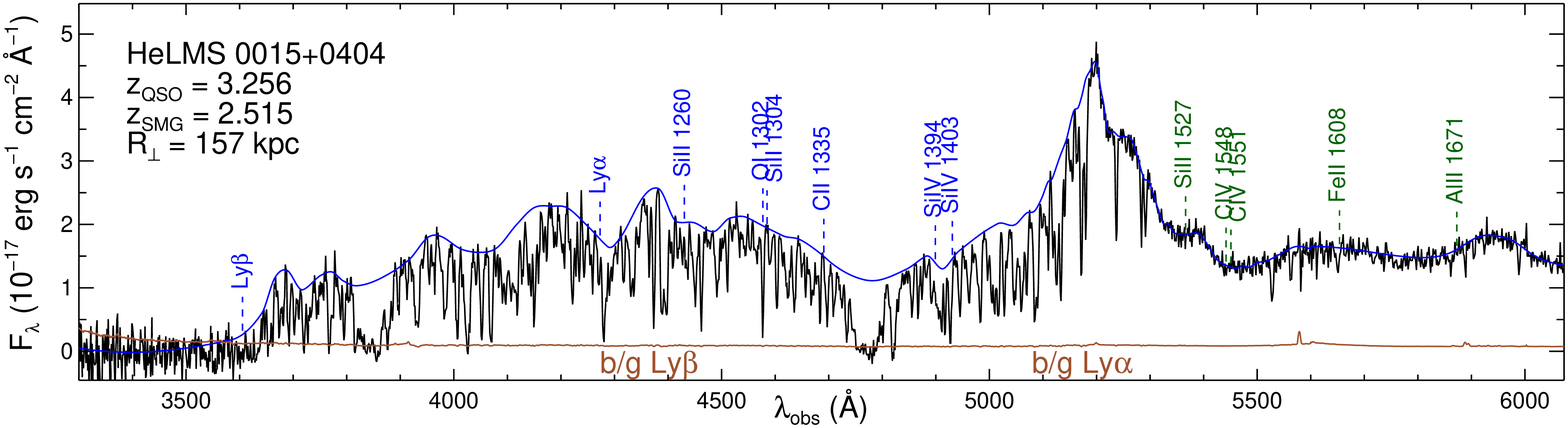}
\plotone{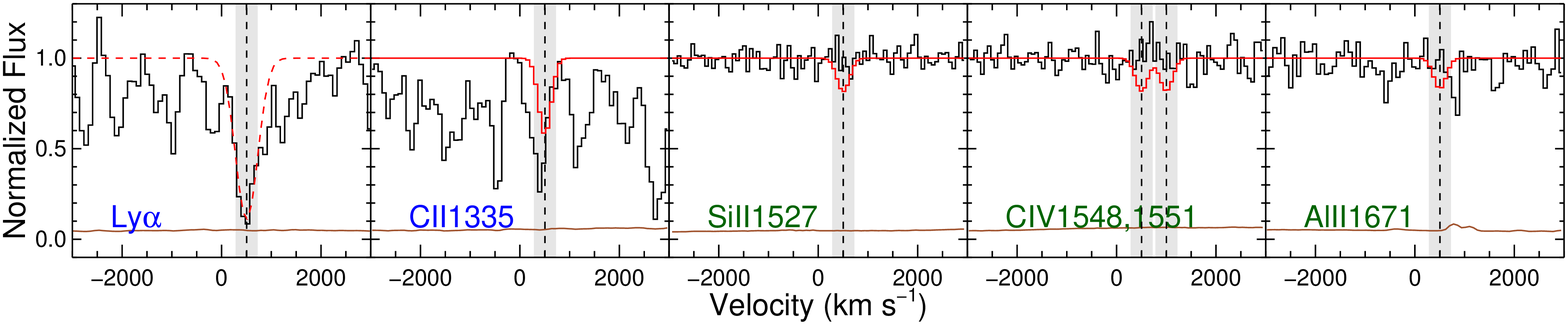}
\plotone{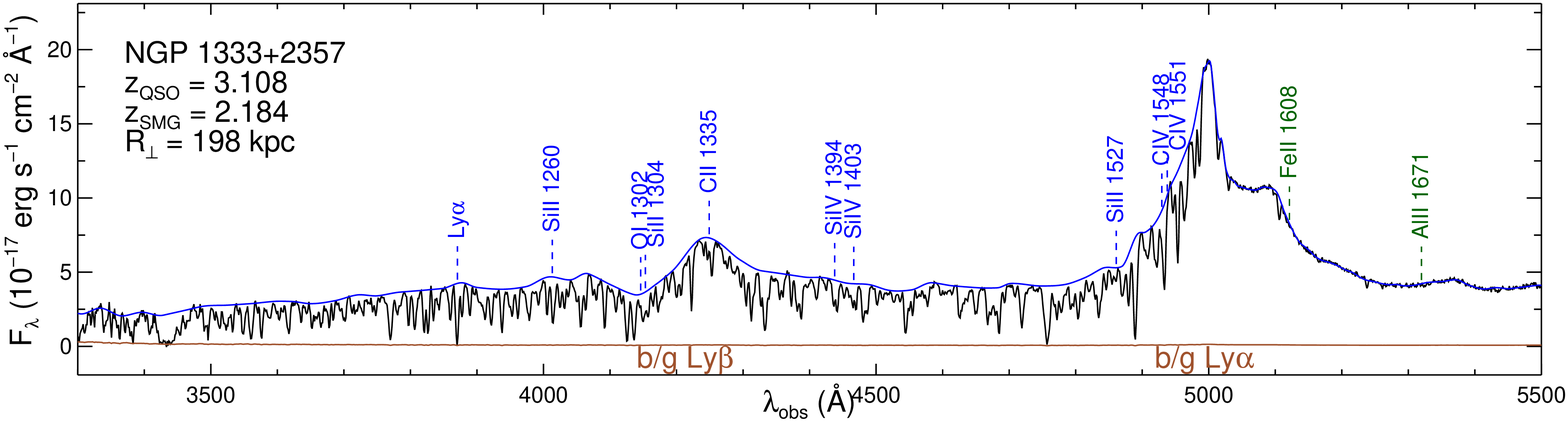}
\plotone{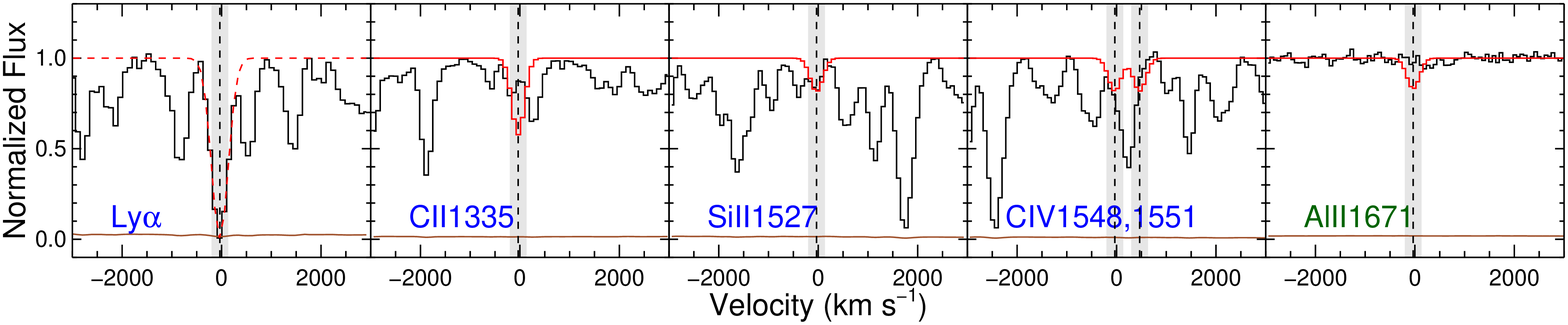}
\plotone{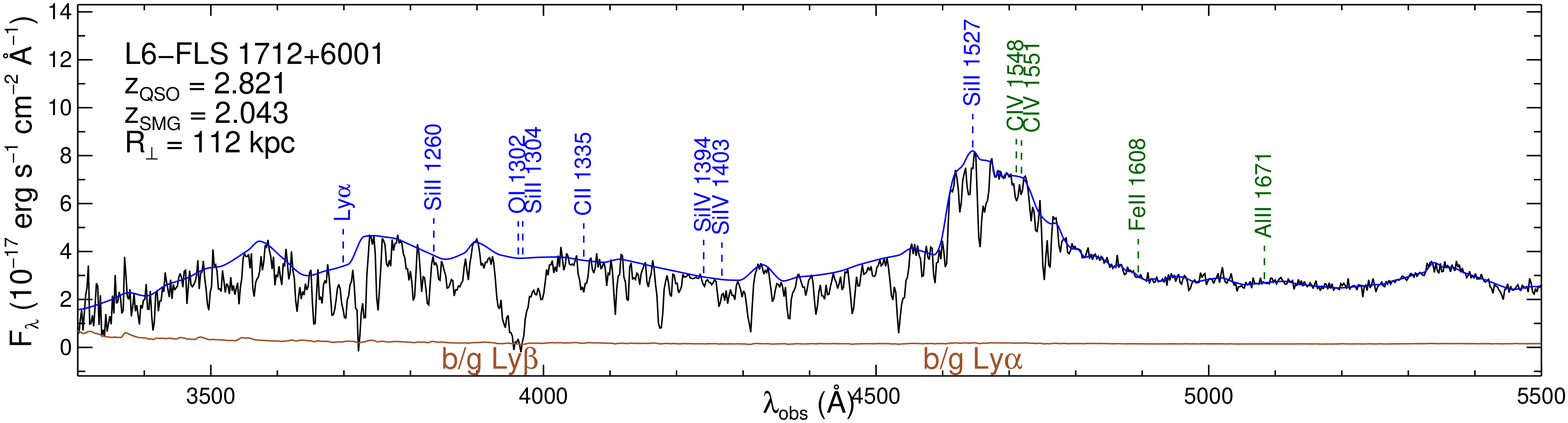}
\plotone{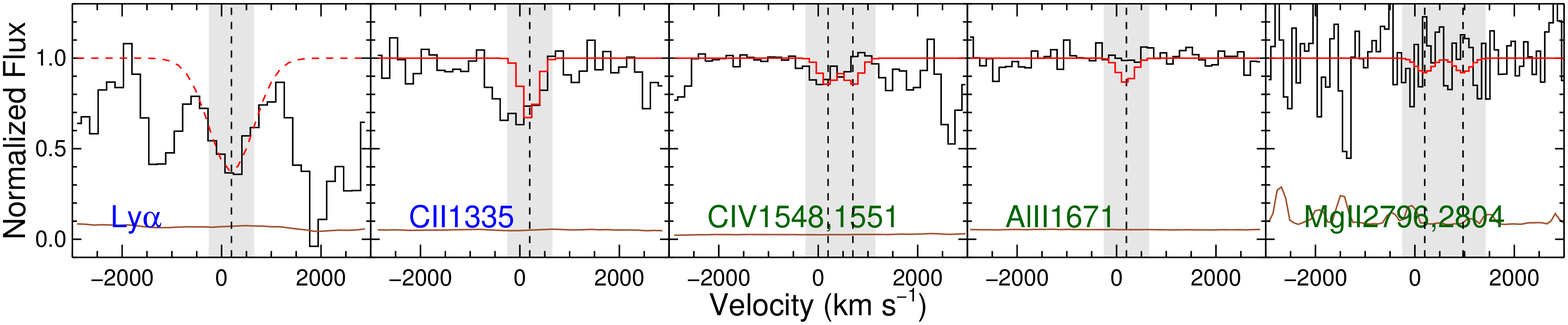}
\caption{Keck optical spectra of background (b/g) QSOs probing foreground SMGs. The black curve shows the observed flux-calibrated spectrum, while the blue curve shows the continuum model used to normalize the spectrum. The expected locations of absorption lines due to the foreground SMG are marked in blue and green for lines inside and outside the Lyman forests, respectively. Below the full spectrum, we also show the velocity profiles for the \HI\ \lya\ and a number of common metal absorption lines. All panels show the region $\pm$3000~\kms\ around the systemic redshifts of the foreground SMGs. In the first panel, we show a Gaussian fit to the strongest \HI\ \lya\ absorption line within our search window of $\pm$600\,\kms. \hl{In other panels, we overlay the metal line profiles of a typical LLS, using the average observed EWs \mbox{\citep[0.6~\AA\ for \CII\ and 0.3~\AA\ for other lines;][]{Prochaska13a}}.} The vertical dashed line shows the centroid velocity and the gray shaded region highlights the $\pm$1$\sigma$ width of the best-fit Gaussian. Both components of the \CIV\ and \MgII\ doublets are highlighted.  
\label{fig:abs}} 
\end{figure*}

\section{Results} \label{sec:result}

\subsection{Radio Detections} \label{sec:radiodetections}

We targeted the \hers\ sources in 13 \sqps\ with the VLA and made six clear detections (46\% detection rate). Figure~\ref{fig:detections} shows the VLA detections and Table~\ref{tab:vlaphoto} lists the 6~GHz positions and flux density measurements for the detections and 3$\sigma$ upper limits on the non-detections. The VLA revealed five genuine \sqps\ (HeLMS\,0015$+$0404, NGP\,1333$+$2357, G15\,1413$+$0058, G15\,1435$+$0110, L6-FLS\,1712$+$6001) and one far-IR-luminous QSO (NGP\,1335$+$2805).  

The VLA-detected SMGs have 6~GHz peak flux densities between 11 and 76~\uJy\ with a mean of $\sim$40~\uJy. For these detections, there is a clear positive correlation between the observed flux densities at 6~GHz and 500~\um, as expected from the IR-radio correlation convolved with a redshift distribution. The 54\% non-detection rate is not surprising given the uncertainties in the predicted 6~GHz flux density and the fact that the sensitivity of radio data may deviate substantially from the theoretical prediction because of confusion sources and weather. 

The offsets between the \hers\ positions and the VLA positions range between 2.7\arcsec\ and 7.9\arcsec. The maximum offset is comparable to the estimated 2$\sigma$ positional uncertainty of \hers\ 250~\um-selected catalogs: $\sigma_{\rm pos} = \sqrt{2} \sigma_{\rm R.A.} = 0.93 \times {\rm FWHM/SNR} = 3.4\arcsec$ for FWHM = 18.1\arcsec\ and SNR = 5 \citep{Ivison07,Smith11}. This indicates that a fraction of the ``pairs'' that have small angular separations ($\theta_{250} \lesssim 8\arcsec$) could be single sources, i.e., QSOs that are far-IR-luminous. In our VLA-detected sample, NGP\,1335$+$2805, with $\theta_{250} = 6.5\arcsec$, is the only such case (see the third panel of Fig.~\ref{fig:detections}). The VLA position agrees with the optical QSO position within 0\farcs1, which is well within the cross-band astrometry accuracy of $\sim$0.3\arcsec\ \citep[SDSS$+$VLA; e.g.,][]{Ivezic02}. But on the other hand, the SMG in L6-FLS 1712$+$6001 is a separate source from the QSO, despite its even smaller 250~\um\ separation ($\theta_{250} = 5.4$\arcsec). We thus chose not to exclude pairs with $5\arcsec < \theta_{250} < 8\arcsec$. The far-IR-luminous QSOs in our sample are interesting on their own right and they will be discussed elsewhere.  

\subsection{Physical Properties of the SMGs} \label{sec:smgproperties}

\begin{deluxetable*}{rccccc ccccc}
\tablewidth{0pt}
\tablecaption{Properties of the Spectroscopically Confirmed \sqps
\label{tab:NHI}}
\tablehead{
\colhead{Pair Name} & \colhead{SMG} & \colhead{$z_{\rm SMG}$} & \colhead{log($L_{\rm IR}$)} & \colhead{$q_{\rm IR}$}  & \colhead{QSO} & \colhead{$z_{\rm QSO}$} & \colhead{$\theta_{\rm 6GHz}$} & 
\colhead{$R_\bot$} &  \colhead{$W_{\rm Ly\alpha}$} & \colhead{\hl{Optical Depth}} \\
\colhead{} & \colhead{(J2000)} & \colhead{}  & \colhead{(\lsun)} & \colhead{} &  \colhead{(J2000)} & \colhead{} & \colhead{(\arcsec)} & \colhead{(kpc)} & \colhead{(\AA)} & \colhead{Classification} 
}
\startdata
HeLMS 0015$+$0404  &J001543.29+040421.4 & 2.515&13.1& 2.1&J001542.31+040433.5 & 3.256& 19.0& 157&$2.0\pm0.2$& ambiguous \\
NGP 1333$+$2357    &J133330.03+235732.7 & 2.184&12.6& 2.3&J133330.36+235709.9 & 3.108& 23.3& 198&$1.7\pm0.2$& ambiguous \\
L6-FLS 1712$+$6001 &J171207.47+600138.0 & 2.043&12.6& 2.6&J171209.00+600144.4 & 2.821& 13.1& 112&$2.9\pm0.5$& optically thin 
\enddata
\end{deluxetable*}

Estimating the intrinsic properties of the VLA-detected \hers\ sources require knowledge of their spectroscopic redshifts. We obtained deep near-IR spectra for the five VLA counterparts whose spectroscopic redshifts are previously unknown. We detected \Ha\ and \NII\ lines from three of the five sources, enabling accurate determination of the spectroscopic redshifts (Fig.~\ref{fig:smgspec} and Table~\ref{tab:NHI}). The \NII/\Ha\ line ratios are consistent with the range observed in typical SMGs ($0.1 \lesssim$~[N\,{\sc ii}]/H$\alpha \lesssim 1.4$; \citealt{Swinbank04}). The two remaining sources, G15\,1413+0058 and G15\,1435+0110, show near-IR continuum emission without detectable emission lines or stellar absorption features. It is possible that the emission lines either fall into one of the telluric absorption bands or are simply outside of the spectral range. Our redshift success rate is thus 60\%, comparable to previous redshift surveys of SMGs in the optical \citep[e.g.,][]{Chapman05, Casey11}. 

With the photometry from \hers\ and the VLA and spectroscopic redshifts, we can estimate the total rest-frame 8$-$1000~\um\ luminosity ($L_{\rm IR}$) and the IR-to-radio luminosity ratio ($q_{\rm IR}$). The results are listed in Table~\ref{tab:NHI}. We fit the \hers\ photometry with a modified blackbody at the spectroscopic redshift, with $\beta$ fixed to 1.5. We find that they have $L_{\rm IR} > 10^{12}$\,\lsun\ and SFR$_{\rm IR}$ = 470-1500~\msunyr. The SFR is estimated from $L_{\rm IR}$ using the calibration of \citet{Murphy11} for a \citet{Kroupa02} initial mass function:
\begin{equation}
 {\rm SFR}/M_{\odot}~{\rm yr}^{-1} = 1.5\times10^{-10}~L_{\rm IR}/L_{\odot}. 
\end{equation} 
The extrapolated 850~\um\ flux densities range between $7 \leq S_{850} \leq 18$~mJy, confirming that they are SMGs. 

The IR-to-radio luminosity/flux ratio is estimated using the following equation \citep{Helou85,Ivison10c}:
\begin{equation}
q_{\rm IR} = {\rm log} \frac{S_{\rm IR}/3.75\times10^{12}~{\rm W}~{\rm m}^{-2}}{S_{\rm 1.4 GHz}/{\rm W}~{\rm m}^{-2}~{\rm Hz}^{-1}}
\end{equation}
where $S_{\rm IR}$ is rest-frame integrated 8-1000~\um\ flux and $S_{\rm 1.4GHz}$ is the rest-frame 1.4~GHz flux density\footnote{This flux ratio is equivalent to the luminosity ratio defined in \citet{Kovacs06}.}. The former is from the modified blackbody fit to the IR SED, and the latter is converted from the observed 6~GHz flux densities assuming a typical synchrotron slope ($S_\nu \propto \nu^{-0.8}$). The three SMGs show $2.1 \leq q_{\rm IR} \leq 2.6$, consistent with the observed radio-IR correlation for \hers\ sources \citep[$q_{\rm IR} = 2.4\pm0.5$;][]{Ivison10c}. This indicates that our sample is not biased to radio-loud AGNs despite being radio selected.

\subsection{Absorption Line Systems}

To compare with previous work on the CGM of other high-redshift galaxies, we need to characterize the covering factor of optically thick gas clouds around SMGs (i.e., Lyman Limit Systems [LLSs] with column densities of neutral hydrogen $N_{\rm HI} > 10^{17.2}$~cm$^{-2}$). We search for \lya\ absorbers near the spectroscopic redshifts of the foreground SMGs and classify the \lya\ absorbers in the QSO spectra following the procedure used in the QPQ study \citep[e.g.,][]{Prochaska13}. \hl{We classify the absorbers as optically thick (i.e., LLSs), optically thin, or ambiguous based on the \mbox{\lya} equivalent widths (EWs) and the presence of associated metal transitions.}

We first identify \HI\ \lya\ absorption due to the SMGs in the QSO spectra within $\pm$600~\kms\ of the systemic redshifts of the SMGs. The search window is chosen because the escape velocity is 610~\kms\ at the virial radius ($R_{\rm vir} = 230$~kpc) of a $10^{13}$~\msun\ dark matter halo at $z = 2$ \citep[][]{Navarro96,Bullock01}. Strong \lya\ absorption lines are detected in all three systems (Fig.~\ref{fig:abs}). We report their rest-frame EWs ($W_{\rm Ly\alpha}$), updated angular separations based on the VLA positions ($\theta_{\rm 6GHz}$), and the impact parameters ($R_\bot$) in Table~\ref{tab:NHI}. As we have found in our earlier QSO absorption line studies, systematic error associated with continuum placement and line blending generally dominates the statistical error of the QSO spectra. So we estimate the error of $W_{\rm Ly\alpha}$ assuming 10\% error in the continuum placement. From our best-fit Gaussians to the absorption profiles (shown in Fig.~\ref{fig:abs}), we find that all three systems show strong \lya\ absorption lines with $W_{\rm Ly\alpha} = 1.7-2.9$~\AA. Note that when $z_{\rm QSO} - z_{\rm SMG} \gtrsim 0.5$, the \HI\ \lya\ absorption from the SMG may lie within the QSO Ly$\beta$ forest. Unfortunately, all three pairs fall in this category, so contamination from \lyb\ lines from systems in the \lya\ forest may be a concern. However, it turned out that \lyb\ contamination is not a serious issue for these systems. Because assuming the detected \lya\ absorption to be \lyb, we searched for the corresponding \lya\ lines in each spectrum but did not find any.

Given the absorption profiles of the \lya\ line, we then searched for the associated metal transitions commonly observed in optically thick absorption systems \citep[e.g.,][]{Prochaska15}: Si\,{\sc ii}\,$\lambda$1260,1304,1527, O\,{\sc i}\,$\lambda$1302, \CII\,$\lambda$1335, Si\,{\sc iv}\,$\lambda$1394,1403, C\,{\sc iv}\,$\lambda$1548,1551, Fe\,{\sc ii}\,$\lambda$1608,2383,2600, Al\,{\sc ii}\,$\lambda$1671, and Mg\,{\sc ii}\,$\lambda$2796,2804. However, none of these metal transitions is convincingly detected in any of our systems. In the following, we discuss the systems individually and provide our classifications:

\begin{itemize}
\item {\it HeLMS\,0015$+$0404}. This system shows an \HI\ \lya\ absorption at $+$500~\kms\ with $W_{\rm Ly\alpha} = 2.0\pm0.2$~\AA\ and {\it intrinsic}\footnote{Instrumental broadening has been deconvolved.} FWHM = 430~\kms\ (Fig.~\ref{fig:abs} {\it top panels}). \hl{The equivalent width gives an upper limit on the \mbox{\HI} column density at log($N_{\rm HI}) \lesssim 18.9\pm0.1$~cm$^{-2}$.} The column density is estimated from the theoretical curve of growth where $W_{\rm Ly\alpha} = 7.3~(N_{\rm HI}/10^{20}~{\rm cm}^{-2})^{0.5}$~\AA\ for $N_{\rm HI} > 10^{18}$~cm$^{-2}$, the regime where the relation is insensitive to the Doppler $b$-parameter \citep[e.g.,][\S16.4.4]{Mo10}. \hl{This is an upper limit because it assumes that the absorption is dominated by a single component, while line blending is likely in the forest.} There is putative \CII\ absorption at $+$400~\kms\ ($W_{\rm CII} = 1.2$~\AA); but it is likely a false identification of a \lya\ forest line, because it is misaligned in velocity with the \lya\ absorption and none of the strong metal lines in the ``clean'' region outside of the Lyman forests is detected (e.g., \SiII, \CIV, \FeII, \AlII). \hl{To assess our sensitivity to these strong metal lines in a typical LLS, we have overlaid the expected absorption profiles on top of the observed spectra in Fig.~\mbox{\ref{fig:abs}}. The model absorption lines are smoothed to the spectral resolution and have the average EWs in LLSs around $z \sim 2$ QSOs: 0.6~\mbox{\AA} for \mbox{\CII} and 0.3~\mbox{\AA} for the other lines \mbox{\citep{Prochaska13a}}. It is clear that the S/N of our spectra are high enough to detect the metal lines from a typical LLS when the lines lie outside of the Lyman forest.} Without detections of strong metal lines and/or the damping wings in \lya, we cannot conclude that the absorber is optically thick based on current data because the high $W_{\rm Ly\alpha}$ could be due to line blending. We conservatively classify this system as ambiguous, even though the non-detection of strong metal lines strongly suggests that it is an optically thin absorber, because fewer than 5\% of optically thick absorbers have no detection of a metal transition \citep{Prochaska15}. 

\item {\it NGP\,1333$+$2357}. This system shows an \HI\ \lya\ absorption at $-34$~\kms\ with $W_{\rm Ly\alpha} = 1.7\pm0.2$~\AA\ and intrinsic FWHM = 260~\kms\ (Fig.~\ref{fig:abs} {\it middle panels}). \hl{Similar to HeLMS\,0015$+$0404, the equivalent width gives an upper limit on the \mbox{\HI} column density at log($N_{\rm HI}) \lesssim 18.8\pm0.1$~cm$^{-2}$.} We observe putative absorption around the expected locations of \CII, \SiIV, \SiII, and \CIV. But all of these lines lie within the \lya\ forest, and their velocity profiles are distinctly different from that of the \lya\ line, indicating that these putative metal absorption are spurious. The \FeII\ and \AlII\ lines are outside of the Lyman forests, but the expected absorption lines are not detected. Like the previous case, we conservatively classify this system as ambiguous.

\item {\it L6-FLS\,1712$+$6001}. There is a broad but shallow \HI\ \lya\ absorption line at $+200$~\kms\ with $W_{\rm Ly\alpha} = 2.9\pm0.5$~\AA\ and intrinsic FWHM = 950~\kms\ (Fig.~\ref{fig:abs} {\it bottom panels}). \hl{The equivalent width gives an upper limit on the \mbox{\HI} column density at log($N_{\rm HI}) \lesssim 19.2\pm0.2$~cm$^{-2}$. But this system is most likely optically thin because the line center drops only to $38\pm4$\% of the continuum intensity (i.e., it does not go line black like the other two systems). Applying the instrumental broadening and pixel sampling to theoretical Voigt profiles, we find that the line center depth places a much stronger limit on the \mbox{\HI} column density than the equivalent width: log($N_{\rm HI}) \lesssim 17.5-15.8$ at $b = 20-30$~\mbox{\kms}, the range of $b$ parameters observed in \mbox{\HI} absorbers at  $z \sim 2-3$ \mbox{\citep[e.g.,][]{Hu95,Rudie12}}.} Like the previous cases, we observe putative metal absorption inside the Lyman forests (e.g., \OI, \CII, \SiII, and \SiIV) but no metal absorption outside of the forests (e.g., \CIV, \FeII, \AlII, and \MgII\footnote{Because of the wider spectral coverage and the lower foreground redshift, this is the only object where we have coverage of the \MgII\,$\lambda$2796,2804 lines.}). Note that the weak \CIV\ absorption is blueshifted by more than 200\,\kms\ from the \lya\ line center. \hl{Considering the strict limit on the \mbox{\HI} column density and the absence of metal lines, we classify this system as optically thin.}

\end{itemize}

In summary, we have identified one clearly optically thin case and two ambiguous cases with three QSO sightlines covering impact parameters between $100 < R_\bot < 200$~kpc around SMGs at $2.0 < z < 2.6$. The two ambiguous cases are also likely to be optically thin, given the non-detection of any metal transitions.

\subsection{Comparison with the CGM around QSOs} \label{sec:discuss}

\begin{figure}[!tb]
\epsscale{1.18}
\plotone{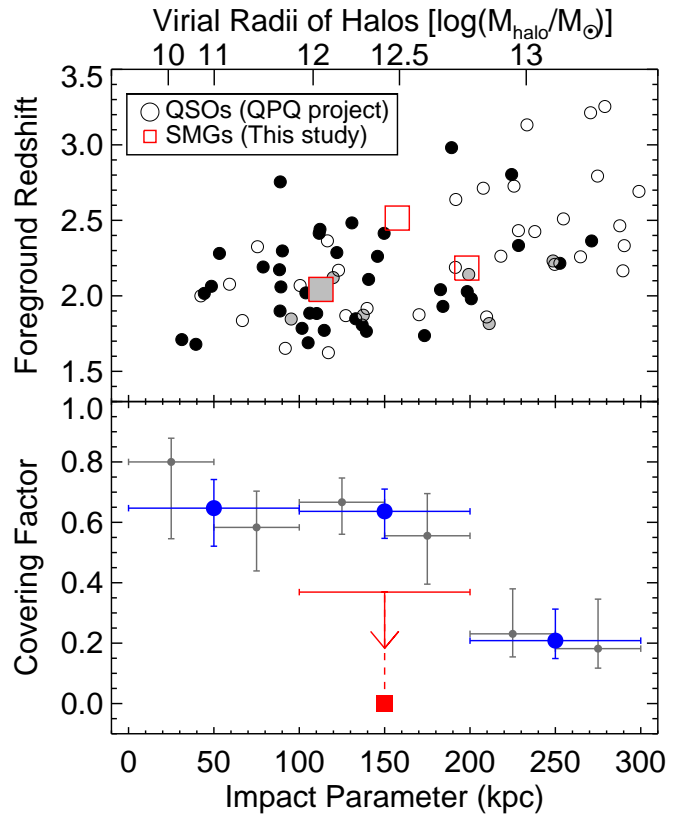}
\caption{
{\it Top:} Sample distributions in foreground redshift vs. impact parameter. The red squares and the black circles show the SMGs from this study and the QSOs from the QPQ survey \citep[][]{Prochaska13}, respectively. The black-filled, gray-filled, and open symbols correspond to systems that are optically thick, optically thin, and ambiguous, respectively. The top axis marks the virial radii for a range of dark matter halos at $z = 2$.
{\it Bottom:} The covering factor of optically thick gas around SMGs ({\it red square with a downward arrow}: our 1$\sigma$ confidence interval for a non-detection in a sample of three) vs. impact parameter, compared to that around QSOs from the QPQ survey \citep[{\it blue/gray circles} with error bars;][]{Prochaska13}. The blue and gray circles are estimates of the covering factor based on 100~kpc-wide bins and 50~kpc-wide bins, respectively. 
Our sample probes similar ranges of redshifts and impact parameters as the QPQ QSOs but shows less optically thick absorbers.
\label{fig:covfrac}} 
\end{figure}

For background QSO sightlines probing the CGM of $z \sim 2-3$ foreground QSOs, one observes \HI\ absorbers optically thick at the Lyman limit $\gtrsim$60\% of the time. The high \HI\ covering factor extends to at least the expected virial radius of $\sim$160~kpc \citep[][]{Hennawi06a,Prochaska13,Prochaska13a}. In Figure~\ref{fig:covfrac}, we compare (1) the SMG sample distribution with the QSOs from the QPQ project, and (2) the covering factor of optically thick \HI\ gas around SMGs with that around QSOs. The two samples overlap in the plane of foreground redshift vs. impact parameter: the SMGs cover the same redshift range as the QSOs in the intermediate impact parameter range between 100 and 200~kpc. We calculate the 1$\sigma$ binomial confidence intervals of the optically thick fraction using the quantiles of the beta distribution \citep{Cameron11}.  Because there is no clearly optically thick absorber among the three systems we analyzed, our 1$\sigma$ confidence interval of the covering factor is 4.2$-$36.9\% for a non-detection in a sample of three. For comparison, we consider all of the QSO sightlines with $100 < R_\bot < 200$~kpc from the QPQ project \citep{Prochaska13} and we find the optically thick covering factor is $64^{+7}_{-9}$\% for 21 clearly optically thick systems among 33 systems. Therefore, despite our small sample, the upper bound of our 1$\sigma$ confidence interval is 3$\sigma$ below the best-estimated covering factor around QSOs at the same ranges of redshifts and impact parameters. 

Note that although our analysis may appear to have limitations based on the classification of two of the SMG absorbers as ambiguous, these same limitations also apply to the QPQ analysis, and are inherent to any attempt to classify absorbers using low or moderate resolution spectra without coverage of the \HI\ Lyman limit (at 912\AA\ in the rest-frame). As such, we have followed exactly the same procedure for absorber classification as in the QPQ studies, enabling a direct comparison to that work. \hl{Our data have shown that clear optically thick cases appear much less frequently around SMGs than around co-eval QSOs. However, if we were to treat the ambiguous cases in both samples as if they were optically thick absorbers, then the difference in the optically thick covering factor becomes smaller because of the high number of ambiguous cases in the SMG sample: the covering factor around SMGs increases to \mbox{$67_{-28}^{+15}$\%} (2 out of 3 systems), while the covering factor around \mbox{$z \sim 2$} QSOs increases to \mbox{$91_{-8}^{+3}$\%} (30 out of 33 systems) between $100 < R_\bot < 200$~kpc.} 

There are some minor differences between our analysis and the QPQ analysis. \hl{But they are unlikely to affect our result.} First, the bulk of the QPQ dataset \citep[][]{Prochaska13a} utilized background QSO spectra with moderate resolution ($R \sim 2000$). This is a factor of two higher than the resolution we used for the QSO spectra probing two of our SMGs ($R \sim 1000$), whereas in L6-FLS\,1712$+$6001 our spectrum has $R \sim 600$. But the lower resolution of our spectra does not play a significant role in the resulting classification of our absorbers, since the typical strength of the metal lines seen in the optically thick systems would still be easily detectable at $R \sim 1000$, and the one system that we observed at $R \sim 600$ is most likely optically thin, since it does not go line black. Another significant difference, is whereas \citet{Prochaska13a} restricted analysis to only those foreground QSOs with redshifts lying within the \lya\ forest of the background QSO, we have included foreground SMGs landing in the \lyb\ forest, owing to the larger redshift separations between the SMGs and the QSOs in our sample. This is unlikely to impact our deduced covering factor in Fig.~\ref{fig:covfrac} though. For example, earlier QPQ studies \citep{Hennawi06,Hennawi07,Hennawi13} also considered absorbers lying in the \lyb\ forest, and the covering factor deduced there was consistent with that derived in \citet{Prochaska13a}. Furthermore, in all of the three \sqps\ shown in Fig.~\ref{fig:abs}, multiple metal lines that are strong in optically thick absorbers (i.e., Si\,{\sc ii}\,$\lambda$1527, C\,{\sc iv}$\lambda\lambda$1548,1551, Al\,{\sc iii}$\lambda$1671, and Mg\,{\sc ii}$\lambda\lambda$2796,2804) land in the clean regions outside of the Lyman forest but remain undetected. Finally, whereas QPQ searched for absorbers within a velocity window in a $\pm$1500\,\kms\ window owing to the large errors in QSO redshifts, we adopted a $\pm$600\,\kms\ search window. We were able to consider a smaller velocity interval because our SMG redshifts derived from narrow rest-frame optical lines are much more accurate, and this interval was instead chosen to encompass the typical virial motions in a $10^{13}$\,\msun\ dark matter halo. The larger velocity interval used in QPQ formally implies a higher level of contamination from physically unassociated absorbers. But the contamination is estimated to be at only $\sim$3\% level for a $\pm$1500\,\kms\ window \citep[][see their Figure 10 and Table 7]{Prochaska13a}. This is too small to be responsible for the difference in the optically thick covering factor that we observe between QSOs and SMGs. 

\section{Summary and Conclusions} \label{sec:summary}

Motivated by the unique properties of SMGs and their purported evolutionary link to high-redshift QSOs and today's massive ellipticals, we have started a project to use QSO absorption line spectroscopy to probe the diffuse cool \HI\ gas in the CGM of SMGs. This work requires a sample of projected \sqps, which are extremely rare. Wide-area sub-millimeter surveys are needed to compile such a sample. Thanks to the advent of \hers, we have identified 163 \sqps\ with bright $z > 2.5$ QSOs ($g_{\rm QSO} < 22$) and angular separations between 5\arcsec\ and 36\arcsec\ from a suite of wide-area \hers\ surveys and spectroscopic QSO surveys. 

Extensive followup observations are required to carry out the absorption line study. 
To allow slit spectroscopy, the first stage is to use an interferometer to pin down the positions of the \hers\ sources. This paper focuses on a subsample of 13 \sqps\ that were observed with the VLA in C-band. With an average integration time of 25~min per source, the VLA detected sources within the \hers\ beam in six fields. One of the six \sqp\ turns out to be a far-IR-luminous QSO at $z = 2.973$, while the \hers\ source in another pair turns out to be a pair of SMGs. 
Hence, we effectively identified six \sqps\ from observations of 13 fields (46\%). 
The second stage is to measure the spectroscopic redshifts of the SMGs. We observed five of the six VLA-identified \sqps\ with near-IR spectroscopy. We were able to determine the redshifts for three of the five SMGs (60\%) from the redshifted H$\alpha$ and \NII\ lines. The remaining two sources show only featureless continuum in the near-IR windows, making it impossible to determine an accurate redshift. 
The last stage is to obtain optical spectroscopy of the QSOs once it is confirmed that the QSOs are in the background of the SMGs. \hl{Because we have selected only the brightest QSOs, the success rate of this step is approaching 100\%}.
Because of the low success rates of the first two followup stages, only $\sim$23\% of the initial sample (i.e., three \sqps\ out of 13) are spectroscopically confirmed and are suitable for final absorption line study. Our main findings are as follows:

\begin{enumerate}

\item The near-IR spectra of the five VLA-detected sources are all detected in hour-long integrations with 8-meter telescopes, although only three of those show emission lines that yielded accurate spectroscopic redshifts. Since we positioned the slits on the VLA positions, this result shows that the spatial offset, if any, between the near-IR and the radio counterparts is less than an arcsec (the slit width), consistent with the finding from differential lensing in strongly lensed sources \citep[e.g.,][]{Fu12b}.

\item The VLA-identified \hers\ ``350~\um peakers'' at $2.0 < z < 2.6$ are similar to SMGs selected at longer wavelengths (i.e., 850~\um\ to 1~mm), in terms of the \NII/\Ha\ ratio (i.e., gas metallicity), the IR luminosity (i.e., SFR), and the IR-to-radio luminosity ratio (i.e., radio excess due to AGN). The VLA-identified SMGs are optically faint and unbiased to radio-loud AGNs, so they indeed represent a galaxy population distinct from the optical selected QSOs and the Lyman break galaxies (LBGs) in the same redshift range. 

\item Strong \HI\ \lya\ absorption is found in the background QSO spectra for all of the three spectroscopically confirmed \sqps\ with impact parameters between $100 < R_\bot < 200$~kpc. Here we have adopted a much narrower search window ($\pm$600~\kms) than the QPQ study, further reducing the level of contamination from physically unrelated clouds. However, none of the three absorption line systems seems optically thick at the Lyman limit (i.e., LLSs with $N_{\rm HI} > 10^{17.2}$~cm$^{-2}$), in contrast to the $\sim$60\% covering factor of LLSs around QSOs from the QPQ study despite similar data quality, foreground redshifts, and impact parameters. 

\end{enumerate}

Our comparison thus suggests {\it either} that SMGs do not have a substantial neutral gas reservoir in their halos that could potentially fuel a prolonged star formation phase {\it or} that SMGs inhabit $\sim10^{12}$\,\msun\ halos so that our sightlines are yet to probe inside their virial radii. If the latter, their halos are comparable to those of LBGs. \citet{Rudie12} found an optically thick covering factor of $30\pm14$\% around LBGs at $z \sim 2.3$ and $R_\bot < 90$~kpc. Note that this is $\sim$2$\times$ lower than that of coeval QSOs and is consistent with the 1$\sigma$ confidence interval that we were able to place for the SMGs. On the other hand, the  difference in the optically thick \HI\ covering factor between SMGs and QSOs casts doubt on the evolutionary link between the two populations, unless AGN outflows can somehow affect the physical state of gas at hundreds of kpc scales within its short lifetime.   

Our final conclusion is limited by the small sample size. To enable a more robust comparison with previous absorption-line studies, we badly need to increase the effective yield of our survey from the current level of $\sim$23\%. In a future publication, we will present observations with the Atacama Large Millimeter/submillimeter Array (ALMA) to pinpoint the positions of the \hers\ sources in the \sqps. Observing at a wavelength (870~\um) much closer to the selection wavelengths (250 to 500~\um), we expect doubling the detection rate with integration times of just several minutes per source. The \hers\ sources in our sample are too faint to allow a quick CO redshift search with ALMA \citep[e.g., the survey of strongly lensed SMGs by][]{Weis13}, but spectrographs on large optical telescopes covering the full optical$+$IR range at a moderate spectral resolution ($R \gtrsim 2000$) will greatly increase the redshift search range and decrease the areas blinded by strong airglow lines.

\acknowledgments

We thank F.~Bian and K.~Allers for providing the data reduction packages, D.~Ludovici for helping with the VLA data calibration of L6-XMM\,0223$-$0605, and the anonymous referee for comments that helped improve the presentation of the paper. 
The National Radio Astronomy Observatory is a facility of the National Science Foundation (NSF) operated under cooperative agreement by Associated Universities, Inc. Support for this work was provided by the NSF through award GSSP SOSPA3-016 from the NRAO. 
H.F. acknowledges support from the NSF grant AST-1614326, the NASA JPL award 1495624, and funds from the University of Iowa.
J.X.P. acknowledges support from the NSF grants AST-1010004, AST-1109452, AST-1109447 and AST-1412981.
A.C. acknowledges support from the NSF grant AST-1313319 and the NASA grants NNX16AF39G and NNX16AF38G.
D.K. acknowledges support from the NSF grant AST-1412153 and funds from the University of California San Diego.
Z.Y.Z. acknowledges support from the European Research Council in the form of the Advanced Investigator Programme, 321302, COSMICISM. 

The \hers-ATLAS is a project with \hers, which is an ESA space
observatory with science instruments provided by European-led
Principal Investigator consortia and with important participation from
NASA. The H-ATLAS website is http://www.h-atlas.org/. The US
participants acknowledge support from the NASA \hers\ Science Center/JPL.
The LBT is an international collaboration among institutions in the United States, Italy and Germany. LBT Corporation partners are: The University of Arizona on behalf of the Arizona university system; Istituto Nazionale di Astrofisica, Italy; LBT Beteiligungsgesellschaft, Germany, representing the Max-Planck Society, the Astrophysical Institute Potsdam, and Heidelberg University; The Ohio State University, and The Research Corporation, on behalf of The University of Notre Dame, University of Minnesota and University of Virginia.
Some of the data presented herein were obtained at the W.M. Keck Observatory, which is operated as a scientific partnership among the California Institute of Technology, the University of California and the National Aeronautics and Space Administration. The Observatory was made possible by the generous financial support of the W.M. Keck Foundation.
Based on observations obtained at the Gemini Observatory, which is operated by the Association of Universities for Research in Astronomy, Inc., under a cooperative agreement with the NSF on behalf of the Gemini partnership: the National Science Foundation (United States), the National Research Council (Canada), CONICYT (Chile), Ministerio de Ciencia, Tecnolog\'{i}a e Innovaci\'{o}n Productiva (Argentina), and Minist\'{e}rio da Ci\^{e}ncia, Tecnologia e Inova\c{c}\~{a}o (Brazil).
The authors wish to recognize and acknowledge the very significant cultural role and reverence that the summit of Mauna Kea has always had within the indigenous Hawaiian community. We are most fortunate to have the opportunity to conduct observations from this mountain.
 
{\it Facilities}: Herschel, Sloan, VLA, Keck/LRIS, LBT/LUCI-1, Gemini/GNIRS


\end{document}